\documentclass[11pt,a4paper]{article}
\usepackage[utf8x]{inputenc}

\usepackage{amsmath,amsfonts,amssymb}

\usepackage{graphicx,graphics}
\graphicspath{{image/}}

\usepackage{cite}
\usepackage[hidelinks]{hyperref}
\usepackage{xcolor}

\usepackage[left=2.5cm,top=2.3cm,right=2.5cm]{geometry}
\usepackage{sidecap}    

\numberwithin{equation}{section}

\title{\Large \textbf{ Islands and Light Gravitons
in type IIB String Theory
\\[5mm]}}

\author{Saskia Demulder$^{\alpha}$\footnote{sademuld@mpp.mpg.de}\ , \quad Alessandra Gnecchi$^{\alpha}$\footnote{agnecchi@mpp.mpg.de}\ , \quad Ioannis Lavdas$^{\beta}$\footnote{ioannis.lavdas@physik.uni-muenchen.de}\ , \quad Dieter L{\"u}st$^{\alpha,\beta}$\footnote{luest@mppmu.mpg.de, luest@theorie.physik.uni-muenchen.de}}
\date{%
\center{$^{\alpha}$\textit{\small Max-Planck-Institut f{\"u}r Physik (Werner-Heisenberg-Institut),\\ F{\"o}hringer Ring 6, 80805, M{\"u}nchen, Germany}}\\
\center{$^{\beta}$\textit{\small Arnold-Sommerfeld-Center for Theoretical Physics,\\ Ludwig-Maximilians-Universit{\"a}t, 80333 M{\"u}nchen, Germany}}\\[2ex]
}

\begin{document}

%%%%%%%%%%%%%%%%%%%%%%%%%%%%%%%%%%%%%%%%
%			TITLEPAGE
%%%%%%%%%%%%%%%%%%%%%%%%%%%%%%%%%%%%%%%%

\begin{flushright}

	\hfill{LMU-ASC 14/22}	

	\hfill{MPP-2022-40}

\end{flushright}

{\let\newpage\relax\maketitle}

\begin{center}
\textbf{Abstract}
\end{center}
\hfill\break

We consider the setup of a black hole in AdS$_{4}$ coupled to an external bath, embedded in type IIB string theory. We study quantum extremal islands in these backgrounds, in relation to the existence of a massive graviton. Using explicit results of the microscopic embedding of AdS$_{4}$ massive gravity  in string theory, we investigate whether it is possible to achieve backgrounds with extremal 
islands, in which the lowest lying graviton is only slightly massive.
For certain regions of the microscopic parameters, the graviton mass can be computed explicitly, and we explain how it directly affects the existence and the properties of the islands.
We also show that islands can in principle exist within the regime of validity of the massive gravity effective field theory.
However we see via numerical computations that the existence of quantum extremal islands at zero temperature is highly constrained, also when the dilaton is allowed to vary, so that the mass of the graviton cannot be made arbitrarily light. At finite temperature, we also identify a critical parameter, above and below which islands still exist but exhibit a different behavior.
Our work supports recent proposals that the unitary evolution of black holes in higher dimensions, and more precisely their Page curve, relies on the presence of a massive graviton in the effective theory.

\thispagestyle{empty}

\newpage
\tableofcontents

%%%%%%%%%%%%%%%%%%%%%%%%%%%%%%%%%%%%%%%%
%			INTRO  ----  Section 1
%%%%%%%%%%%%%%%%%%%%%%%%%%%%%%%%%%%%%%%%

\section{Introduction}

Recent progress in the study of systems of black holes coupled to an external bath has lead to important developments in deciphering the evolution of information during black hole evaporation\footnote{This has fueled an intense research activity over the last couple of years, which has been captured by recent reviews and progress reports \cite{Almheiri:2020cfm,Raju:2020smc,Bousso:2022ntt}.}, reproducing the Page curve of an evaporating black hole \cite{Penington:2019npb,Almheiri:2019psf,Almheiri:2019hni}. Advances in this direction rely on a deeper understanding of the calculation of the entropy of Hawking radiation, which has been possible thanks to the identification of a new contribution to the entanglement entropy, following from a quantum generalization \cite{Engelhardt:2014gca} of the holographic entanglement entropy prescription \cite{Ryu:2006bv,Hubeny:2007xt}. This novel contribution is due to \emph{islands}, quantum extremal surfaces physically disconnected from the asymptotic radiation region, that dominate the entropy contribution after the Page time, ensuring a unitary evolution of the system \cite{Engelhardt:2014gca,Penington:2019npb,Almheiri:2019psf,Almheiri:2019hni}. The role of quantum extremal islands in reproducing the Page curve of black hole evaporation has been extensively studied in models of lower dimensional gravity, where it corresponds to the contribution of wormhole geometries to the gravity path integral \cite{Penington:2019kki,Almheiri:2019qdq}. Not withstanding that methods to directly compute the entanglement entropy of quantum fields in a conformal field theory (CFT) are best developed in lower dimensional systems, the extension to higher dimensional setups is one of the most significant problems that has emerged from the islands prescription.

Proposals to investigate islands in higher dimensions have appeared \cite{Almheiri:2019psy,Geng:2020qvw,Geng:2020fxl,Geng:2021hlu,Geng:2021mic}, based on gravitational setups that include a Randall-Sundrum brane \cite{Randall:1999vf,Karch:2000ct,Karch:2000gx}. These are models of localized gravity, which we will refer to as Karch-Randall setups, in which a 4d end-of-the-world (EOW) brane is added to the anti-de-Sitter AdS$_5$ bulk. The brane ends on the asymptotic region of the AdS$_5$ space, cutting the boundary off at the intersection. A black hole living on the brane will evaporate in a part of the AdS$_5$ boundary, the radiation region. In this framework, the entanglement entropy of quanta in the radiation region can be computed holographically via the Ryu-Takayanagi prescription: the contribution of quantum extremal islands is captured through the identification of a corresponding extremal surface in the AdS$_5$ bulk. 

The Karch-Randall branes setup can be interpreted as gravitational realizations of dual boundary CFTs in presence of a defect \cite{Aharony:2003qf}. Holographic string theory configurations realizing conformal field theories with boundary and interfaces are known, and can be constructed via explicit uplifts to, e.g., type IIB string theory \cite{DHoker:2007zhm,DHoker:2007hhe,Assel:2011xz,Aharony:2011yc}. In this context, quantum extremal islands and their contribution to the black hole entropy have been studied in \cite{Uhlemann:2021nhu}, making more precise the relation of these systems to Karch-Randall configurations. Also bottom-up models of interface CFTs have been investigated recently in \cite{Suzuki:2022xwv,Anous:2022wqh}, which clarify the correspondence between the gravitational system coupled to a CFT and the BCFT, and derive general results for entaglement entropy in braneworld setups.

The main physical lessons emerging from higher dimensional studies of quantum extremal islands are that 1) when the black hole is coupled to a non-gravitating bath,  there must be a massive graviton in the theory for islands contribution to become dominant at some stage in black hole evaporation\cite{Geng:2020qvw}, and 2) islands extremal surfaces undergo a phase transition related to the variation of the underlying geometric parameters (brane tension, ratios of the number of branes in the microscopic setup...) \cite{Geng:2020qvw,Uhlemann:2021nhu}.

In this paper, we will address the study of quantum extremal islands in a family of the type IIB backgrounds. In particular we will consider the uplift of a Janus-like geometry to type IIB string theory, extending the analysis of \cite{Uhlemann:2021nhu} both presenting a quantitative discussion on the graviton mass in the low energy theory, as well as considering more general backgrounds with varying dilaton. In this analysis the black hole is coupled to a non-gravitating bath\footnote{In the case of a gravitating bath, previous works have consistently showed that the Page curve is not correctly reproduced in Karch-Randall setups or in type IIB uplifts \cite{Laddha:2020kvp,Geng:2020fxl,Uhlemann:2021nhu}.}. Crucially, in a particular regime of these solution one disposes of a \textit{quantitative} control over the mass of the graviton. Indeed, by looking at  a specific range of microscopic parameters, through the study of the spin two wave equation it is possible to derive a quantitative expression for the lightest spin 2 field \cite{Bachas:2011xa,Bachas:2018zmb}, both for non-varying dilaton as well as for background where $\delta\phi\neq0$. This will enable us to clarify the role of the mass of the graviton on the appearance of islands in this particular context.

We will investigate the properties of island surfaces both for empty Anti de Sitter (\emph{zero temperature}) and when the black hole is present (\emph{finite temperature}). 
 We will see that islands at zero temperature are bound to a region of parameter space,  where  the graviton mass is above a certain critical value. This critical value must be contrasted with the natural ultraviolet (UV) cutoff \cite{deRham:2016plk} of a massive gravity effective field theory (EFT), and we will discuss the relevant regime of the microscopic parameters to ensure that the range of validity of the EFT does not exclude the presence of islands. Notice that the embedding of this EFT in the UV has been provided in \cite{Bachas:2018zmb}, so we will not have to worry about the validity of the massive gravity EFT which, in general, can be subject to Swampland constraints \cite{Klaewer:2018yxi,DeRham:2018bgz}.

After ensuring the consistency of the EFT with the existence of islands, we will study the properties of these surfaces in backgrounds with varying dilaton.

Our main result is that increasing the dilaton variation affects the microscopic parameter range, within which the islands still contribute (at zero temperature), or they undergo a phase transition (at finite temperature). This directly reflects on the mass of the graviton, which depends both on the microscopic parameters and the dilaton variation. More precisely, we find that, when allowing the dilaton to vary, one can find overlapping regimes where the graviton is only slightly massive and at the same time islands contribute at zero temperature. However, islands cannot be found for arbitrarily large dilaton variation $\delta\phi$, where the graviton mass can tuned to be arbitrarily low. This shows that the above is not possible, if one also wants to preserve the properties of quantum extremal islands: the latter enter in fact a critical regime as the graviton mass is increased. Furthermore, this yields another confirmation that, in higher dimensions, islands and black hole evaporation are physically related to the graviton mass.

The present work is structured as follows. In Section 2 we review the Karch-Randall setup of a black hole coupled to a thermal bath, as well as the realization of such configuration in IIB string theory via an explicit microscopic uplift. We consider in particular backgrounds with varying dilaton, and we present the expression for the mass of the graviton.
In Section 3 we discuss the regime of validity of the EFT below the cutoff $\Lambda_{EFT}$ of \cite{deRham:2016plk}, and how this translates into a bound on the number of D-branes of the 10 dimensional setup. Section 4 is devoted to the detailed analysis of islands in background with varying dilaton. Our results are discussed in comparison to the constant-dilaton setup \cite{Uhlemann:2021nhu}, which one retrieves in the $\delta\phi=0$ limit of our analysis. We end the paper with conclusions in Section 5, where we stress once again how our results provide evidence that the island contribution to black hole entropy is related to the presence of a graviton with mass above a critical value.

%%%%%%%%%%%%%%%%%%%%%%%%%%%%%%%%%%%%%%%%
%			REVIEW ON ISLANDS  -- Section 2
%%%%%%%%%%%%%%%%%%%%%%%%%%%%%%%%%%%%%%%%

\section{Entanglement islands in higher dimensions: From braneworld models to type IIB embeddings}
\label{sec:review}

In the present section, we summarize the main attributes of the basis of this work, namely the braneworld models describing AdS black holes coupled to an external, non-gravitating bath, and their realization via string theory uplifts, consisting on a class of warped AdS$_{4}$ backgrounds. We will deal with special limits of the internal manifold, in which the lowest-lying graviton has a parametrically small mass, and we also briefly review of the corresponding massive gravity embedding.

\subsection{Karch-Randall braneworlds and Page curve}

\begin{figure}[h!] 
\centering
\includegraphics[width=0.5\textwidth]{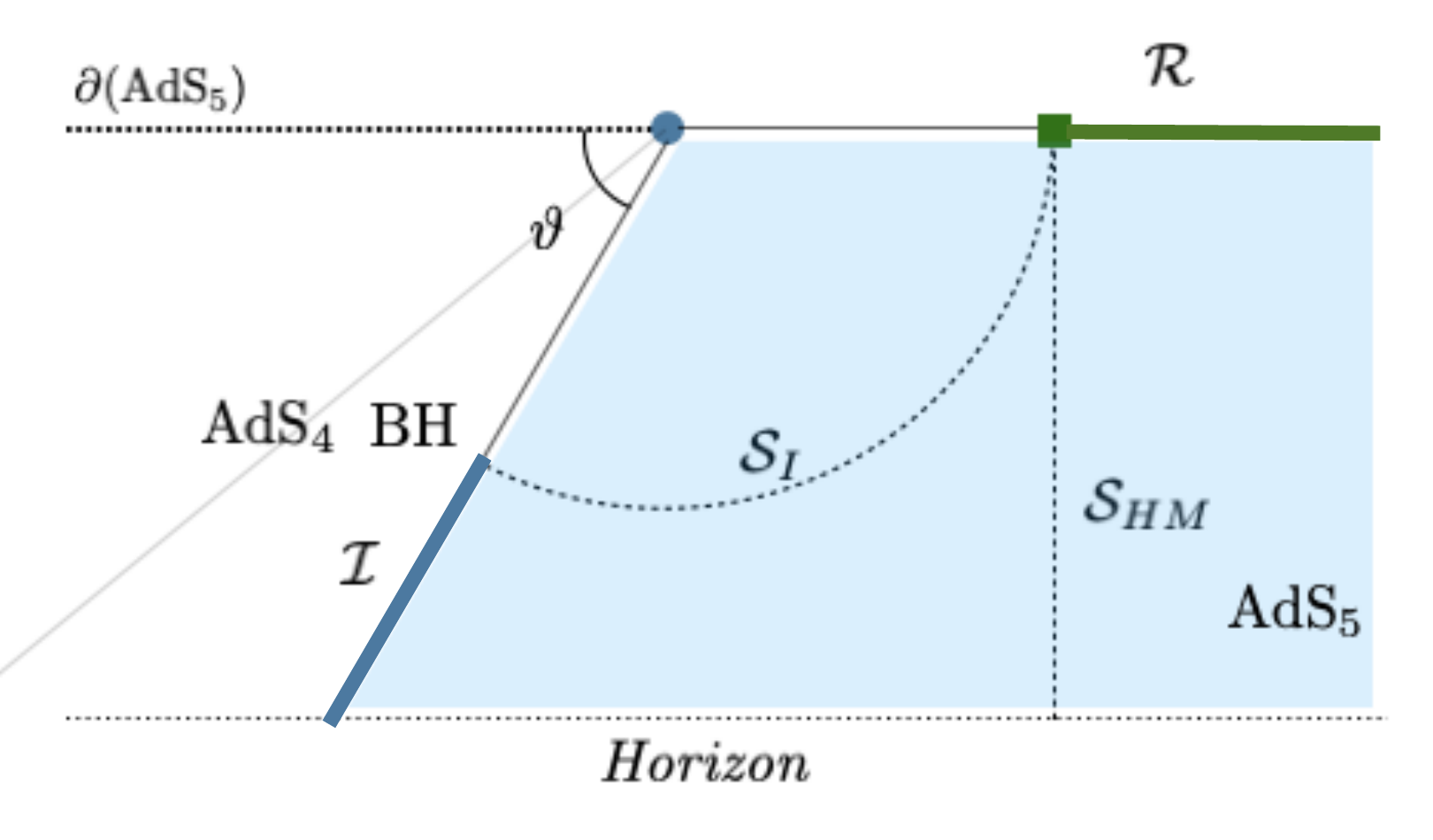}
\caption{\small{Karch-Randall setup: funnel configuration with AdS$_4$ black hole on the EOW brane within the AdS$_5$, bent at an angle with respect to the conformal boundary of the bulk. The blue segment on the brane is the Island, on the boundary of which is anchored the island surface, ending on the endpoint of the radiation region, denoted by the green segment,  on the conformal boundary. The dashed line segments correspond to the two kinds of RT surfaces. }}\label{kr}
\end{figure}

The most promising realization of quantum extremal islands for evaporating black holes in higher dimensions is the Karch-Randall (KR) braneworlds \cite{Randall:1999vf, Karch:2000ct}. They offer in fact important insights on the calculation of the entanglement entropy of emitted radiation, thanks to their doubly holographic nature: these configurations can be seen in fact either as $d$-dimensional conformal field theories (CFT$_{d}$) on half-space coupled to a $(d-1)$-dimensional boundary theory, or a gravitational system on the AdS$_{d+1}$ bulk sliced by an end of the world (EOW) AdS$_{d}$ brane, or CFT$_{d}$ on the half space and AdS$_{d}$ coupled to CFT$_{d}$ on the brane. These three description are related by holographic duality, applied to either the full system, or just to the boundary of AdS$_{d+1}$. The last description, depicted in Fig. \ref{kr}, is central for the study for the black hole/bath system. 

The present work focuses on a four dimensional funnel configuration \footnote{Recall that only in the funnel solution the Hawking radiation can be encoded in the classical 5d bulk, as opposed to the droplet solutions, for details see section 3 of \cite{Geng:2020qvw} as well as \cite{Geng:2020fxl}. }, consisting of a black-hole AdS$_{4}$-brane embedded in an AdS$_{5}$ bulk and  intersecting its conformal boundary. This setup provides a standard configuration of an evaporating black hole coupled to an external, non-gravitating bath, in which two kinds of entanglement extremal surfaces appear that contribute to the Page curve: the Hartman-Maldacena surface ($\mathcal{S}_{\text{HM}}$) and the Quantum Extremal Island surface ($\mathcal{S}_{I}$) \cite{Geng:2020qvw, Geng:2020fxl}. At early times, the Page curve is characterized by a linear growth reflecting the dominance of the $\mathcal{S}_{\text{HM}}$ over $\mathcal{S}_{I}$, while after some characteristic time the latter dominates, and the entropy saturates. This is due to the fact that the area of the $\mathcal{S}_{\text{HM}}$ increases with time while the  area of the $\mathcal{S}_{I}$ remains constant. 

In the braneworld constructions, a relevant parameter characterizing the configuration is the angle $\vartheta$ formed between the EOW brane and the boundary. Its significance, at zero temperature, relies on the fact that it introduces a regime of contribution of the islands to the generalized entanglement entropy.  It has been found in fact that island surfaces in empty Anti de Sitter roughly disappear once the angle falls below a certain value, referred to as the critical angle $\vartheta_{c}$.  On the other hand, this behavior is regulated by the horizon in the AdS black string background, and islands contribute also for angles below the critical one \cite{Geng:2020fxl}. Understanding how island surfaces are related to a critical parameter in the type IIB uplift corresponding to this braneworld scenarios is one of the main points we will address in this work.

 The graviton is massive due to the coupling of the brane to the bath, as a result of transparent boundary conditions. The next section will be entirely devoted to the analysis of the consequences a massive graviton has in the EFT where islands are supposed to contribute to the entanglement entropy. What we would like to review here is that from doubly holographic models, it can be observed that the massless graviton limit corresponds to the vanishing angle limit $\vartheta\to 0$ where the EOW-brane approaches the conformal  boundary of AdS$_{5}$. In the massless limit, the islands cease to contribute. This can be understood from the behavior of the island surface in this limit, both endpoints of which now approach $\partial\text{AdS}_{5}$. The surface anchors on the boundary of the radiation region and on the boundary of the island, as seen below. In the limit  where $\vartheta\to 0$, this surface falls into the Poincaré horizon. The divergence of the surface is explained by the divergence of the inverse of the gravitational constant $G$ on the brane \cite{Geng:2020qvw}.

\subsection{IIB string theory embeddings}\label{sec:IIBembedd}
Having reviewed the braneworld constructions, in this section we will introduce the family of corresponding type IIB embeddings that will be the protagonist of our study of quantum extremal islands.

\paragraph{Uplifting 4d/5d configurations} Karch-Randall double holographic models, as CFTs ending on a boundary, can be actually explicitly embedded in IIB string theory. The embedding is described by the warped background $AdS_{4}\times_{w}\mathcal{M}_{6}$, where the six dimensional internal manifold is a warped product of two-spheres over a Riemann surface. When the internal manifold $\mathcal{M}_{6}$ is compact, these backgrounds are holographically dual to 3d $\mathcal{N}=4$ super conformal field theories (SCFT) engineered in type IIB with D5, NS5 and D3 branes \cite{Assel:2011xz,Aharony:2011yc}. In the case where the internal manifold $\mathcal{M}_{6}$ is non-compact, the geometry asymptotically becomes AdS$_5\times$S$^5$ and the solutions are dual to a 3d defect SCFT in 4d $\mathcal{N}=4$ super Yang-Mills. The 10d geometry is parametrized by a pair of harmonic functions, with logarithmic singularities at points on the boundary of the Riemann surface, representing the locations of D5 and NS5 branes. The AdS$_{4}$ fiber degenerates at these points, with one of the two $S^{2}$'s shrinking to a point at each singularity. Non-contractible three and five-cycles support non-vanishing three-form and five-form fluxes, indicating that the background can be seen as the near horizon limit of intersecting IIB brane configurations, where D3 branes are suspended between five-branes. 

In this work we consider a particular case of the general setups introduced in \cite{Assel:2011xz}. The left asymptotic region smoothly caps off and is substituted by an interior point of the full ten-dimensional space while the geometry asymptotes to global $AdS_{5}\times S^{5}$ on the right\footnote{Strictly speaking, these backgrounds correspond to a realization of a ``fat" Karch-Randall brane, given by the NS5-D5-D3 bound state, while the original KR setup uses a thin brane; both descriptions model the same defect CFT configuration \cite{Bachas:2018zmb}.}. The resulting background is given by the following metric:
\begin{equation}\label{metric}
ds_{10}^{2}=L_{4}^{2}ds_{\text{AdS}_{4}}^{2}+ f_1^{2}ds_{S_1^{2}}+{f_2}^{2}ds_{{S_2}^{2}}+4\rho^{2}dz d\bar{z},
\end{equation}
where $z$ and $\bar z$, parametrize the Riemann surface:
    \begin{align}
        \Sigma&=\left\{ z=x+iy\ \large|\ x\in \mathbb{R} \,, \ y\in\left[0,\frac{\pi}{2}\right] \right\} \ ,
    \end{align}
the warp factors are
	\begin{align}
		L_4^8&=16 \frac{N_1 N_2}{W^2}\ ,
		\qquad 
		\rho^8 = \frac{N_1N_2 W^2}{h^4 \hat{h}^4} \ ,
		\qquad
		f_1^8=16h ^8\frac{N_2 W^2}{N_1^3} \ ,
		\qquad
		f_2^8 =16 \hat{h}^8 \frac{N_1 W^2}{N_2^3} \ ,
	\end{align}
and 
\begin{gather}
	\begin{aligned} 
		W&= \partial h\bar\partial \hat{h} + \bar\partial h\partial \hat{h} = \partial\bar\partial (h \hat{h}) \ ,
		\\
		N_1&=2h\hat{h}|\partial h|^2-h^2 W \ ,
	    \\
		N_2&=2h\hat{h}|\partial \hat{h}|^2-\hat{h}^2 W \ .
	\end{aligned}
\end{gather}
Finally, these functions are determined by two harmonic functions which, in our setup, will take the form\footnote{In their standard form, the harmonic functions are labelled by a set of parameters controlling the asymptotic regions of the geometry, the variation of the dilaton between the asymptotic regions, the D5 brane charges and finally the position of the point singularities on the boundaries of the strip. For a complete presentation and details of the supergravity solutions, the reader is referred to \cite{DHoker:2007zhm, DHoker:2007hhe, Assel:2011xz}. } (we set $\alpha'=1$):
\begin{gather}
	\begin{aligned} \label{eq:har-fun}
	% this would have index 1 or no hat:
	h&=-\frac{i\pi}{4}e^{z}\kappa-\frac{N}{4}~\text{log}\,\text{th}\Big(\frac{i\pi}{4}-\frac{z}{2}\Big)+\text{c.c}\, ,\\
	% this would have index 2 or a hat:
	\hat{h}&=\frac{\pi}{4}e^{z}\hat\kappa-\frac{N}{4}~\text{log}\,\text{th}\Big(\frac{z}{2}\Big)+\text{c.c}\, .
	\end{aligned}
 \end{gather}
This setup will generalize the one of  \cite{Uhlemann:2021nhu}, by considering a nonzero dilaton variation in the $AdS_5$ throat as in \cite{Bachas:2018zmb}, given by
	\begin{align}\label{eq:vardilkappa}
		e^{2\delta\phi}=\frac{\hat\kappa}{\kappa} \ .
	\end{align}
The configuration with a black hole is obtained by substituting $ds^2_{AdS_4}$ with the metric of a black hole in AdS$_{4}$. Since the geometry considered here is capped off on the left, the black hole will be radiating in the right asymptotic region, playing the role of a non-gravitating bath. 

\begin{figure}[h!] 
\centering
\includegraphics[width=0.8\textwidth]{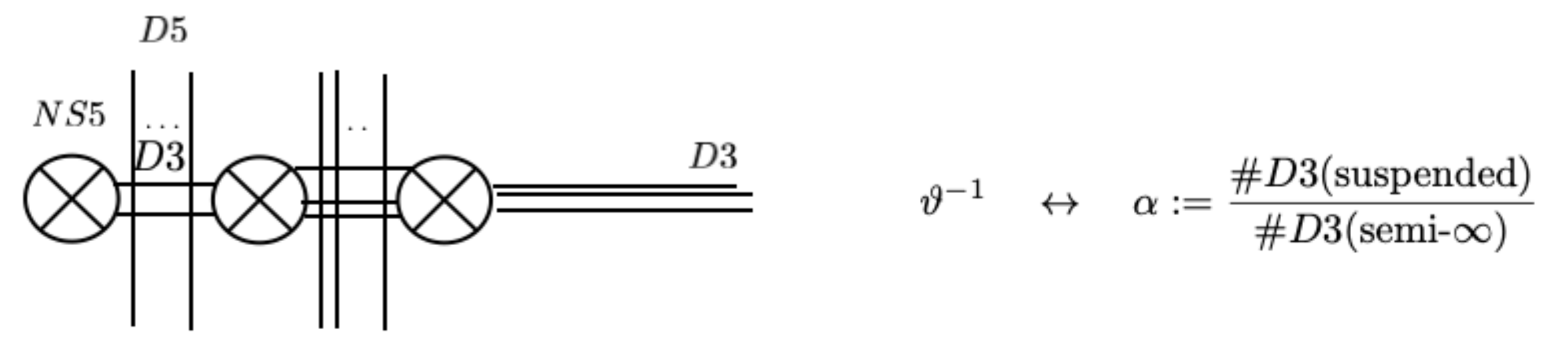}
\caption{\small{The geometric ratio $\alpha$ corresponds to the relative number of D3 branes that are either semi-infinite or suspended between the D5 branes in the brane configuration realizing the IIB embedding with a gravitating AdS bath.}}\label{fig:GeoRatio}
\end{figure}

The solution described by the harmonic functions in \eqref{eq:har-fun} includes two point singularities localized on the strip boundaries at $(0,\frac{\pi}{2})$ and $(0,0)$. For $\kappa=\hat\kappa\equiv K$ there are N-D5 and N-NS5 branes respectively, and $2NK$ semi-infinite D3 branes along the asymptotic region on the right. In \cite{Uhlemann:2021nhu} it was shown how the extremal island surfaces are crucially dependent on the parameter $\alpha= N/ K$, which is the corresponding critical parameter of the KR setups. This parameter will be recovered in our setup as\footnote{If we define 
$\kappa=K e^{-\delta\phi}$, and $\hat\kappa = K e^{\delta\phi}$, 
then $\alpha=N/K$ which holds for generic $\delta\phi$, and reduces to the parameter of \cite{Uhlemann:2021nhu} when $\delta\phi=0$.
}:
\begin{align}\label{alphadef}
	\alpha = \sqrt{\frac{N^2}{\kappa \hat\kappa}} \ .
\end{align}
As presented above, the angle $\vartheta$ between the EOW brane and the conformal boundary of AdS$_{5}$ determines at zero temperature the contribution of the islands to the entropy. In the string theory embedding at hand, the quantity corresponding to the angle $\vartheta$  of the lower dimensional setups,  is the ratio $\alpha$ of the D3 branes suspended between the fivebranes over the semi infinite D3s of the solution, see Fig. \ref{fig:GeoRatio}. Practically this quantity is a ratio of ``boundary'' over ``bulk'' degrees of freedom and can be interpreted as the ``tension'' of the composite brane (NS5-D5-D3 system). Therefore, for large values of $\alpha$ the angle $\vartheta$ becomes small and vice versa: $\alpha \leftrightarrow 1/\vartheta$. The behavior of islands crucially changes upon variation of parameters above or below $\alpha_{\text{crit.}}$. 

\noindent
The analysis of  \cite{Uhlemann:2021nhu} indicates that at zero temperature, there exists a critical value of this ratio below which the island surfaces dominate over the Hartmann-Maldacena surface and the islands contribute to the entropy leading to a Page curve.

In practice determining this critical value $\alpha_{\text{crit.}} $ is based on numerical analysis that studies the divergence of the distance between the anchoring points of the island surface on the black hole region and on the external bath. The numerical computation of the critical ratio will be explained and performed in Section \ref{sec:numerics}. This behavior is analoguous to the low-dimensional one where with respect to critical value of the angle parameter, subcritical setups are characterized by island contribution and hence by Page curve-like behavior of the entanglement entropy. 

At this point, we underline the fact that the existence of a critical value for the geometric ratio $\alpha$ for which the islands stop to contribute is an attribute of the zero temperature solution, which is an uplift of the corresponding empty-AdS braneworld model presented in the previous subsection. On the contrary, as in the AdS-black string background, at finite temperature this effect is regulated by the black hole horizon and islands continue to contribute even for values of the geometric ratio higher than the critical one determined by the zero temperature solution. 

A central point to be noted is that in the above solution the lowest-lying graviton is massive, as a result of the non-compactness of the internal manifold \cite{Bachas:2011xa}. Nevertheless an explicit computation of its mass is only possible on the special limit of the geometry where the mass is parametrically small.

\paragraph{Light gravitons}

The type IIB string embedding corresponding to the metric \eqref{metric} is a consistent uplift of the corresponding braneworld models. In this construction, the lowest-lying spin-2 mode acquires a mass due to the fact that the internal manifold is non-compact.  In the following we review how this type IIB background can be appropriately adjusted in order to allow for such a nearly vanishing graviton mass, which can then be computed explicitly. After briefly presenting the corresponding setup we will continue by examining, using the results of the previous section, whether islands contribute in the limit where the graviton mass is nearly non-zero.

\begin{figure}[h!] 
\centering
\includegraphics[width=0.5\textwidth]{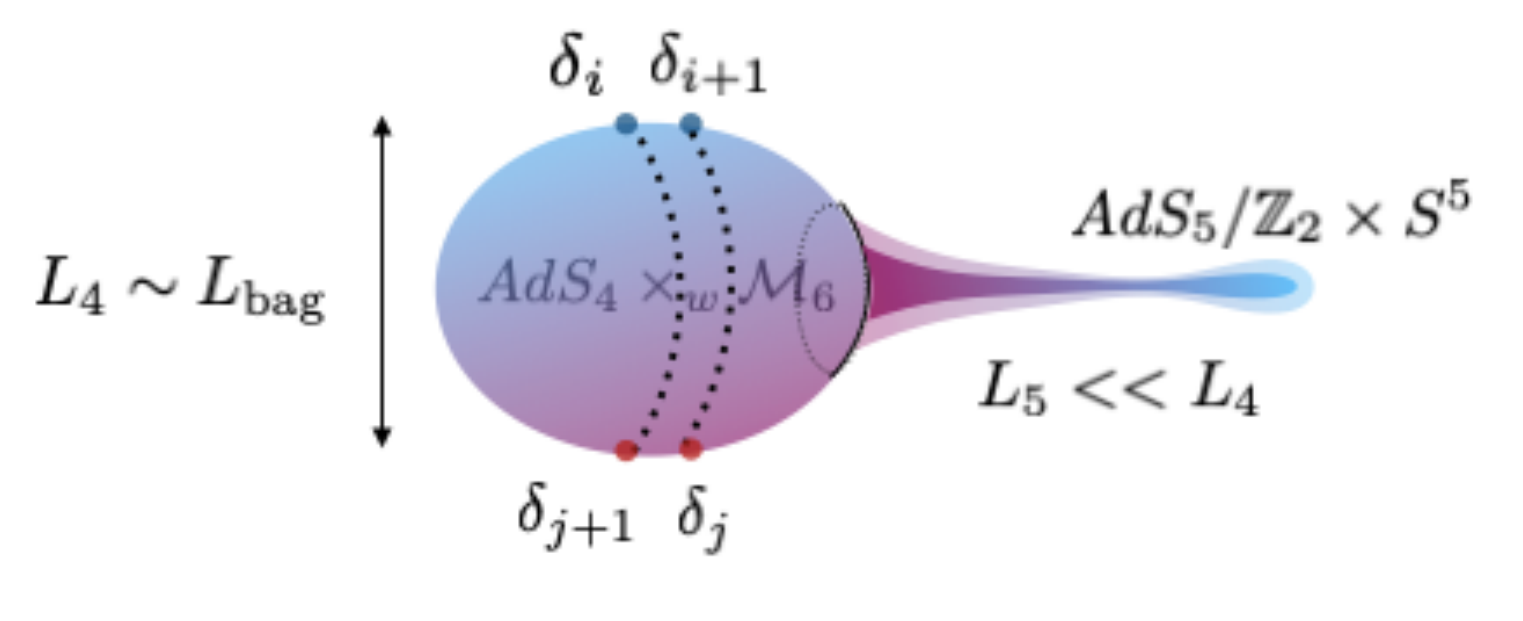}
\caption{\small The geometry of interest: The compact bag is attached to a semi-infinite throat with radius hierarchically smaller than the size of the six-dimensional internal manifold. The fivebranes are found on the point singularities located at $\delta, \hat{\delta}$.}\label{One-bag-pinching}
\end{figure}

 The AdS$_{4}$ vacuum where the internal manifold $\mathcal{M}_{6}$ is compact, includes a massless low-lying graviton and a KK tower of massive modes. The geometry in which the graviton acquires a mass,  includes on the contrary a non-compact internal manifold.  The spin-2 spectrum can be extracted numerically by the corresponding eigenvalue problem for the Laplace-Beltrami operator on the internal manifold \cite{Bachas:2011xa}. Choosing to focus instead on the lowest lying spin-2 mode one solves the variational problem giving the minimal eigenvalue of the mass squared operator \cite{Bachas:2018zmb}.  The background allowing for a small mass for the lowest-lying spin-2  consists of a compact region in the vicinity of the fivebrane singularities (\textit{bag}) attached to a semi-infinite AdS$_{5}\times$S$^{5}$ throat under the constraint that the radius of the throat is much smaller than the size of the bag\footnote{It has been proven recently that the graviton mass is very light in generic compactifications consisting of a small throat connecting two compact geometries \cite{DeLuca:2021ojx}.} ($L_{5}<< L_{\text{bag}}\sim L_{4}$). This can be seen in the Fig. \ref{One-bag-pinching}, where $L_{5}$ is the throat radius and $L_{4}$ the AdS$_{4}$ radius, bound to the bag-size $L_{\text{bag}}$ as a result of the scale non-separation. 

The reason for this can be well understood holographically \cite{Bachas:2018zmb}: Consider a CFT$_{3}$, dual to the AdS$_{4}$ solution above. This theory has a conserved stress-tensor $T_{(3)}^{\mu\nu}$, with canonical conformal dimension, as a result of the conservation. Coupling this three-dimensional theory to a four dimensional $\mathcal{N}=4$ super Yang-Mills in the bulk, results to the violation of the conservation of the stress tensor. Since the two theories are coupled, the three-dimensional stress tensor is no longer conserved but rather leaks in the bulk direction. In particular, we consider that the stress tensor is dissipating weakly in the extra bulk direction, a fact that implies that its (canonical) dimension acquires a small anomalous dimension.  Holographically, the anomalous dimension of the stress tensor corresponds to a small correction to the $AdS_{4}$ graviton mass:

\begin{equation}\label{eq:holo-dissip}
m_{g}^{2}L_{4}^{2}=\Delta(\Delta-3)\xrightarrow{\Delta=3+\varepsilon}{}m_{g}^{2}L_{4}^{2}\sim \varepsilon
\end{equation}

The weak dissipation is ensured by the fact that the number of degrees of freedom of the boundary theory is much greater than the one of the bulk theory, hence
	\begin{align}
		m^2_g L_4^{2} \ \sim \  \frac{\text{bulk dof}}{\text{bdy dof}} \ .
	\end{align}
In terms of the dual solutions, the scarcity of the bulk degrees of freedom correspond to the small D3 brane number in the semi infinite throat as opposed to the large number of D3 branes connecting the D5 and NS5 branes in the bag-region. The graviton mass, obtained by solving the variational problem for the minimal mass eigenvalue corresponding to the graviton wavefunction can be written explicitly as \cite{Bachas:2018zmb}:
\smallskip
\begin{equation}\label{eq:grav_mass}
m_{g}^{2}L_{4}^2\equiv \bar{m}_g^2\simeq \frac{3\pi^{3}}{4}\Big(\frac{L_{5}}{L_{\text{bag}}}\Big)^{8}\mathcal{J}(\text{ch}(\delta\varphi))
\end{equation}
\smallskip
at leading order in the radii ratio. The last factor gives the contribution of the dilaton to the graviton mass\footnote{In the case of varying dilaton the background develops a Janus throat, a one-parameter deformation of $AdS_5\times S^5$ geometry \cite{DHoker:2007zhm}.}: if it does not vary through the throat, $\mathcal{J}(\text{ch}(\delta\varphi))$ reduces the graviton mass for increasing dilaton variation, while its contribution becomes trivial in the case of a constant dilaton. As we will see in the following analysis, the dilaton variation plays a central role in the discussion of the island surfaces. In the case of the solution at hand, parametrized by \eqref{eq:har-fun}, the radii are given by $L_{5}^{4}\sim NK$ and $L_{\text{bag}}^{4}\sim N^{2}$ , with the expression for the mass taking the form: 

\begin{equation}\label{eq:grav_mass_alpha}
m_{g}^{2}L_{4}^2\equiv \bar{m}_g^2\simeq \frac{3\pi^{3}}{4}\Big(\frac{K}{N}\Big)^{2}\mathcal{J}(\text{ch}(\delta\varphi)).
\end{equation}

\begin{figure}[ht]
\centering
\includegraphics[width=0.55\textwidth]{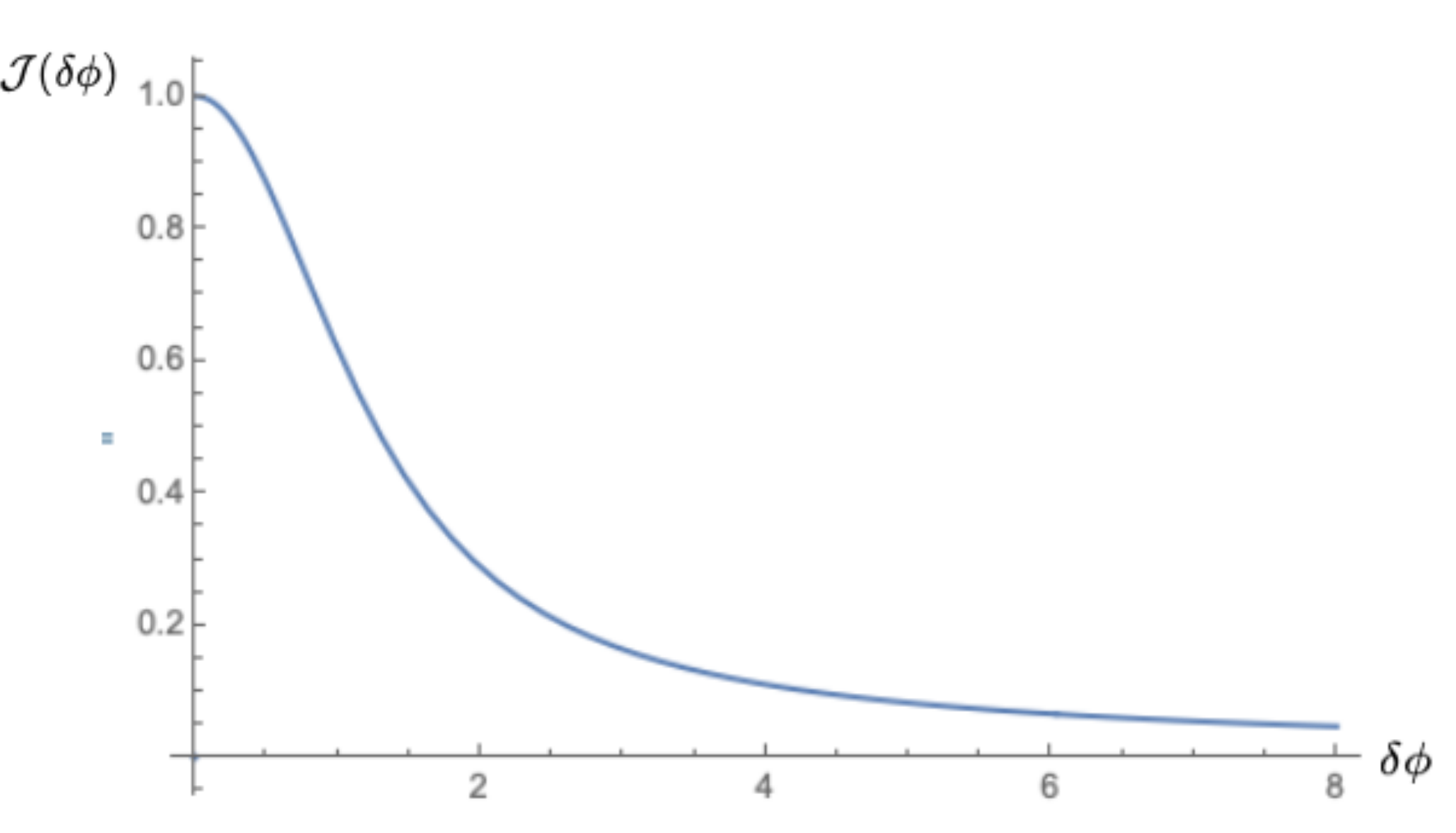}
\caption{\small The Janus correction function. Observe that for vanishing dilaton variation the mass is not affected.}
\end{figure}

The above background is realized as a special limit of the bimetric gravity embedding \cite{Bachas:2017rch}, where the throat connects two compact regions, with the spectrum including also a massless graviton, while it is related under an appropriate deformation to solutions with multiple massive gravitons \cite{Lavdas:2020tyd}.

\noindent
In this work, by studying the uplift of the 4d/5d embedding, we analyse how to tune  the parameters of the geometry in such a way to allow for a slightly massive graviton and study properties of the island surfaces under this modification of the embedding. Finally, we focus on the effect of the dilaton variation on the behavior of the island surfaces and on the condition for reconciling small graviton mass and island contribution.

%%%%%%%%%%%%%%%%%%%%%%%%%%%%%%%%%%%%%%%%
%			GRAVITON MASS ---- Section 3
%%%%%%%%%%%%%%%%%%%%%%%%%%%%%%%%%%%%%%%%

\section{Massive graviton, islands and EFT validity regime}\label{sec:EFT}

Let us start from the simple configuration of Section \ref{sec:IIBembedd}, with $N$-D5 branes connected by D3 branes to $N$-NS5 branes, residing in the upper and lower point singularities respectively.  The background is engineered so that the dilaton is constant $\delta\phi=0$. We are going to discuss, in this section, the graviton mass and the EFT cutoff in relation to the critical value of the geometric parameter, $\alpha_{\text{crit.}}$, characterizing the island surfaces. 

When the background develops a small-radius throat connecting the AdS$_4$ to the AdS$_5$ geometry, as explained in the previous section and depicted in Fig. \ref{One-bag-pinching}, the background has a slightly massive graviton, whose mass is given from \eqref{eq:grav_mass_alpha}, which for $\delta\phi=0$ takes the form:

\begin{equation} \label{mgbar-alpha}
\bar{m}_{g}^{2}\simeq\frac{3\pi^{3}}{4}\frac{1}{\alpha^{2}} \ .
\end{equation}

The dependence of the mass on the parameter $\alpha$ is a feature expected from the analysis in Section 2 and the holographic interpretation of the graviton mass, as we already saw in eq. \eqref{eq:holo-dissip}: the mass of the graviton decreases for increasing number of  bag-degrees of freedom, and hence for increasing value of the parameter $\alpha$.
Since the parameter $\alpha=N/K$ is the corresponding quantity of $\vartheta^{-1}$ in the KR setup, this implies that the graviton mass is also directly related to the angle: a small graviton mass $\bar{m}_{g}^{2}$ corresponds to small values of $\vartheta$ and vice versa. Finally,  in the KR setups island surfaces are not expected to contribute for values $\vartheta< \vartheta_{\text{crit.}}$ or equivalently in 10d for $\alpha>\alpha_{\text{crit.}}$. This means that for a background with fixed bag degrees of freedom (namely with fixed number of five-branes, $N$), islands are expected to contribute where the graviton mass is above some critical value:

\begin{equation} \label{grav-crit-mass-constraint1}
\bar{m}_{g}>\bar{m}_{g,\text{crit.}}\sim\frac{1}{\alpha_{\text{crit.}}}
\qquad\leftrightarrow \qquad
{K>K_{\text{crit.}}} \ .
\end{equation}

For a given 10d background,  the value of $\alpha_{\text{crit.}}$ can be determined numerically. However, for consistency with small graviton mass, from eq. \eqref{mgbar-alpha}, $\alpha_{\text{crit.}}$ must be be sufficiently large in this regime (this translates into $N\gg K$). For models where $\alpha_{\text{crit.}}$ is not found to be sufficiently large to ensure a hierarchy between $L_4$ and $L_5$, (or equivalently between the defect and the bulk CFT degrees of freedom) the mass formula \eqref{eq:grav_mass} is not valid.

\begin{figure}[ht] 
	\centering 
	\includegraphics[width=0.5\textwidth]{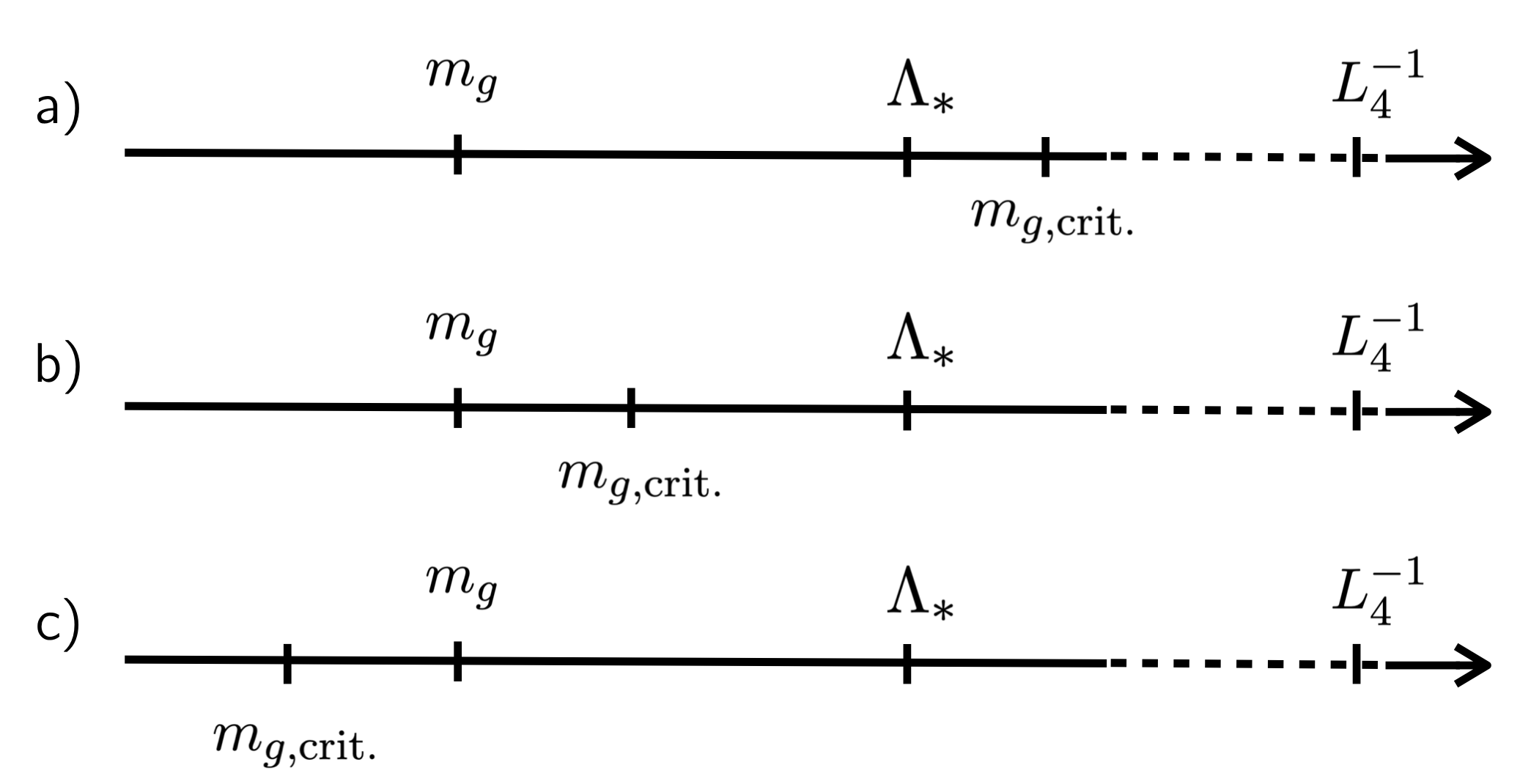}
	\caption{\small For the range $m_{g\,\text{crit.}}<m_{g}<\Lambda_{*}$, shown in case c), the island contributes to the entropy while the theory has a valid EFT description. } \label{EFT-vs-mg}
	\end{figure}

For an AdS massive gravity theory in the regime $m_g\ll L_4^{-1}$ (i.e. $K\ll N$), the EFT breaks down at a scale given by \cite{deRham:2016plk,Bachas:2019rfq} 
\begin{equation}\label{EFT-cutoff}
	\Lambda_{*}=\frac{m_{g}^{\frac{1}{3}}M_{\text{Pl}}^{\frac{1}{3}}}{L_{(4)}^{\frac{1}{3}}}\, ,
\end{equation}
where it is already implied that $m_g < \Lambda_*$, in an AdS background with fixed $L_4$ ($M_{\text{Pl}}$ is the 4d Planck Mass that will be set to 1). Then, the critical values of the graviton mass for the island surfaces can be outside (case a)) or inside the regime of validity of the EFT (case b) and c) of Fig. \ref{EFT-vs-mg}). For the critical mass to be an energy scale at least below the EFT cutoff, we must require
	\begin{align}
		m_{g\,\text{crit.}} < \Lambda_*
		 \ \ \Rightarrow \ \ 
		K_{\text{crit.}}< K^{1/3} N^{5/6} \ll N^{7/6}\ ,
	\end{align}
where we have averaged\footnote{From non-separation of scales, recall that : $\langle L_{(4)}^{2}\rangle_{\text{bag}}\sim L_{\text{bag}}^{2}\sim N$. Averaging over the 6d internal manifold: $ \langle f\rangle_{\text{bag}}=\int \sqrt{g}f/\sqrt{g}$.} the expression on ${\cal M}_6$. Recall that, due to the no-scale separation, $\langle L_{(4)}^{\frac{1}{2}}\rangle_{\text{bag}}\sim L_{\text{bag}}^{\frac{1}{2}}\sim N^{\frac{1}{4}}$.

A more stringent regime corresponds to the requirement that the graviton mass is above the critical mass value, which simply translates into
	\begin{align} \label{constr-mg-above-mgcrit}
		m_{g\,\text{crit.}} < m_g 
		\ \ \Rightarrow \ \ 
		K_{\text{crit.}} < K \ .
	\end{align}
Notice that the first bound varies if one varies the AdS$_4$ vacuum, while the second case is independent on the scale of the Anti de Sitter curvature $L_4$ of the background under consideration. Only if we require \eqref{constr-mg-above-mgcrit} islands always exist in the massive gravity EFT, for consistency with \eqref{grav-crit-mass-constraint1}. 
These bounds reproduce once again the fact that islands contribution to entanglement entropy are in tension with a parametrically small graviton mass.

We can compare the results of this general analysis with the numerical study of \cite{Uhlemann:2021nhu} for backgrounds of the form considered in this section. There, a value of $\alpha_{\text{crit.}} \sim {\cal O}(1)$ was found, which is way below the value needed to reach a parametrically small graviton mass. The rest of the paper will be dedicated to an analysis of a more general background by introducing a varying dilaton through the harmonic functions $h$ and $\hat h$ with two different parameters $\kappa\neq\hat\kappa$, to investigate how the presence of a $\delta\phi\neq0$ affects the critical value of the microscopic parameter $\alpha$. 

Let us conclude this section with a comment on the $m_g \to 0$ limit. In principle, in Karch-Randall double holographic models, one can smoothly and continuously tune the angle $\vartheta$ to zero. From the string theory microscopic embedding, this cannot happen since the critical parameter $\alpha$ depends on the number of D3 branes, which is quantized. 
Another way to tune the graviton mass to zero is to take $\delta\phi\to +\infty$, as can be seen from eq. \eqref{eq:grav_mass}, which corresponds to a decoupling of the bulk theory. However, this limit is singular and can be seen as an infinite distance limit in moduli space \cite{Bachas:2018zmb}, showing that the decoupled theory is no more consistent with the 10d uplift: reaching that point in the dilaton moduli space would bring down an infinite tower of modes rendering the EFT invalid. In this limit, the compactness of the geometry is restored and the graviton becomes massless, so we can interpret it as the limit of the gravitating bath regime. In that case, however, the system has a flat Page curve \cite{Laddha:2020kvp,Uhlemann:2021nhu}, in line with the findings of \cite{Bachas:2019rfq} that one cannot smoothly connect the non-compact background with the compact one, that would correspond to the $\delta\phi\to +\infty$ limit.

%%%%%%%%%%%%%%%%%%%%%%%%%%%%%%%%%%%%%%%%
%	  RESULTS WITH VARYING DILATON ----- Section 4 
%%%%%%%%%%%%%%%%%%%%%%%%%%%%%%%%%%%%%%%%

\section{Island surfaces and running dilaton}\label{sec:numerics}

In this section we study entanglement islands in the type IIB solution introduced in section \ref{sec:IIBembedd}, namely we consider the aforementioned 10d background with a dilaton that is allowed to vary between the bag $AdS_4\times{\cal M}_6$ region and the $AdS_5\times S^5$ bath. We investigate how this variation affects the island surfaces. To this aim, we perform a numerical study of the surface PDE from which we extract the critical value $\alpha_{\text{crit.}}$, which characterizes the island surfaces behavior. We discuss our findings in relation to the possibility of being in a regime with sufficiently small graviton mass, either in order to trust the massive gravity description from perturbative string calculations \cite{Bachas:2018zmb} or to stay within the valid EFT regime, as we discussed in Section \ref{sec:EFT}.

Notice that the massless graviton limit corresponds to the limit of large dilaton variation across the throat (at fixed geometric ratio $\alpha$), from \eqref{eq:grav_mass}. We have already mentioned that this decoupling limit is singular, thus we expect that the extremal islands will be affected by a large dilaton variation, as we confirm with our numerical analysis, both at zero and at finite temperature.

In our numerical analysis we do not solve for other extremal surfaces, like Hartmann-Maldacena surfaces, which we expect to behave similarly to the zero dilaton case, derived in \cite{Uhlemann:2021nhu}.

\subsection{Critical parameters and graviton mass}

\begin{figure}[h!]
\centering
\includegraphics[width=0.6\textwidth]{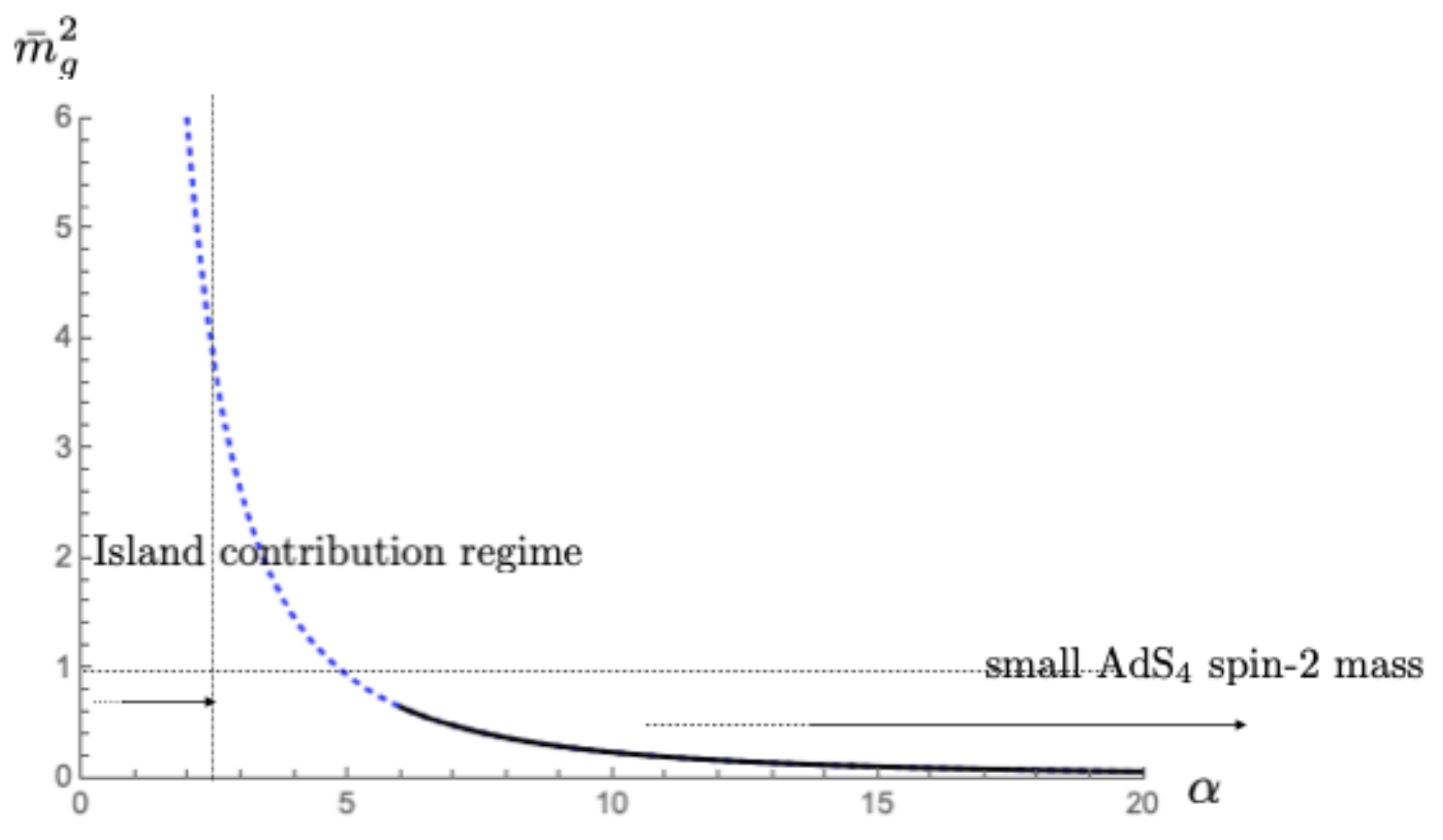}
\caption{\small Zero temperature configuration: it can be observed that parametrically small masses $(\bar{m}_{g}^{2}\ll 1)$ are achieved in the regime of $\alpha\gg 5.0$.  Only the black part of the curve can be computed using the expression for the mass.}\label{fig:graph_mass_vs_alpha_zeroTemp}
\end{figure}

The massive gravity regime which is under control in string theory requires the graviton mass to be $\bar{m}_g^2\ll 1$. It would be ideal to have a numerical evidence that puts non-critical islands in the same regime, but the previous results of \cite{Uhlemann:2021nhu} have found the critical value of the geometric parameter\footnote{We thank C. Uhlemann for pointing out a technical issue in the numerical analysis in an earlier version of this paper.} $\alpha_{\text{crit.}} \sim 4$ . From \eqref{eq:grav_mass} we see that this value is not large enough to create a small throat between the AdS$_4$ and AdS$_5$ region, thus we are in the regime $m_{g\,,\text{crit.}} \sim L_4^{-1}$, which is outside of the validity regime of the EFT: parametrically small graviton masses are achieved for values of the geometric ratio $\alpha$ which lie way above its critical value (see e.g. figure \ref{fig:graph_mass_vs_alpha_zeroTemp}).  In this section, we will see how the critical value changes once the dilaton is allowed to vary in a setup analogous to the one of \cite{Uhlemann:2021nhu}, and how one can enter the regime of small graviton masses.

By introducing a varying dilaton, the graviton mass not only depends on $\alpha$, but also on $\delta\phi$. If the properties of island surfaces physically depended on $\alpha$ only, and not on $\delta\phi$, we would be able to choose a background with islands below the critical value $\alpha_{\text{crit.}}$, and then increase the dilaton variation such that the graviton mass becomes only slightly nonzero. This would be in contrast with expectations that the islands are strongly related to the presence of a massive graviton, and in fact what we find is that it is the combination of $\alpha$ and $\delta\phi$, thus the graviton mass, which effectively governs the island behavior.

\subsection{Numerical setup for island surfaces at $\delta\phi\neq0$}

We consider backgrounds with running dilaton introduced in Section \ref{sec:IIBembedd}. The dilaton value in the bag region is determined by the number of five-branes, while in the bath $AdS_5\times S^5$ region its value is a free parameter determining the Yang-Mills coupling constant of the dual theory. The geometry is specified by the harmonic functions \eqref{eq:har-fun}, and the dilaton variation \eqref{eq:vardilkappa}.

The island surfaces in this family of type IIB backgrounds are minimal surfaces that extend from the bag region to the AdS$_5$ bath. As in \cite{Uhlemann:2021nhu}, the minimal surfaces are eight dimensional surfaces embedded in the ten dimensional geometry. The surfaces completely wrap the two $S^2$'s, and partially wrap both the Riemann surface $\Sigma$ and the AdS$_4$ black hole, whose geometry is parametrized as 
\begin{align*}
	\mathrm ds_4^2=\frac{\mathrm d r^2}{b(r)} + e^{2r}(-b(r) \mathrm dt^2 + \mathrm ds^2_{\mathbb R^2})\,,
\end{align*}
with $b(r)=1-e^{3(r_{h}-r)}$, where $r_{h}$ is the horizon radius while $b(r)=1$ being the zero temperature limit. We take $r_{h}=0$ throughout our numerical analysis.

The extremal surfaces are thus realized as embedding surfaces in terms of the AdS$_4$ coordinate of the black hole geometry, $r(x,y)$ where $x\in \mathbb R$ and $y \in [0, \pi/2]$ are the real and imaginary part of the coordinates on the Riemann surface $\Sigma$. The extremality condition for this surface can be expressed in terms of the functions characterizing the harmonic functions \eqref{eq:har-fun} pinpointing the type IIB solution described in Fig. \ref{One-bag-pinching}:
\begin{align*}
	f=|h \hat h W|\,, \quad g=\frac{1}{2b(r)}\left|\frac{h \hat h}{W} \right|\, ,
\end{align*}
where $W=\partial_z\partial_{\bar z}(h\hat h)$, which yield the differential equation \cite{Uhlemann:2021nhu}
\begin{align}\label{eq:min_cond}
\frac{1}{1+g(\nabla r)^2}\left[2-\nabla 
(g\nabla r)+\frac{1}{2}g \nabla r\cdot \nabla \ln \left(\frac{1+g(\nabla r)^2}{b(r)f^2} \right) \right]=0	\,,
\end{align}
here $\nabla$ is the covariant derivative on $\Sigma$.   The minimal surface is obtained by letting the extremality condition \eqref{eq:min_cond} stabilise after waiting for a sufficiently long `relaxation time' $\tau$
\begin{gather}
\begin{aligned}\label{eq:min_cond_tau}
	\partial_\tau r(x,y,\tau)%&= -L_\gamma ^{-1}\frac{\delta L_\gamma}{\delta r(x,y,\tau)} \,,\\
&=-\frac{1}{1+g(\nabla r)^2}\left[2-\nabla (g\nabla r)+\frac{1}{2}g \nabla r\cdot \nabla \ln \left(\frac{1+g(\nabla r)^2}{b(r)f^2} \right) \right]\,,
\end{aligned}
\end{gather}
where now $r$ is not only a function of $x,y$ but also of $\tau$ (though not $f$ and $g$). The accuracy of the numerical solution at a fixed relaxation time $\tau_\mathrm{max}$  can be measured by the residual
\begin{align}\label{eq:residual}
	R=\left|\frac{1}{1+g(\nabla r)^2}\left[2-\nabla (g\nabla r)+\frac{1}{2}g \nabla r\cdot \nabla \ln \left(\frac{1+g(\nabla r)^2}{b(r)f^2} \right) \right]\right|_{\tau=\tau_\mathrm{max}}\,.
\end{align}
The problem is completely defined when the boundary conditions are specified. The Janus deformation of the throat geometry does not affect the original problem and boundary conditions considered in \cite{Uhlemann:2021nhu}, which we recall here for completeness. In the bag region, the surfaces are smooth in the limit $x\rightarrow -\infty$ which corresponds to a regular point of the 10d geometry. At the singular points, located on the two boundaries of the Riemann surface (strip) the surfaces fall into the horizon. The surfaces continue in the bath and where the right $AdS_{5}\times S^{5}$ asymptotic emerges, they anchor on the bath at a specific positive value of the radial coordinate. This translates into a Dirichlet boundary condition at $x\rightarrow +\infty$ where the embedding function $r(x,y)$ has a constant value, the anchoring point $r_R$:
\begin{equation}
\lim_{x\to +\infty}r(x,y)=r_{R} \ .
\end{equation}
This boundary condition is not affected by the dilaton variation, as the surface still anchors on $AdS_{5}\times S^{5}$, where there is no $y$-dependence and most importantly the dilaton has a constant value\footnote{The $\delta\phi$ dependence concerns in fact the throat region and not the bag, nor the radiation region.}.  

Summarizing, the boundary condition for the minimal surface are captured by 
\begin{align}
	\partial_yr(x,y)\mid_{y=0,\frac{\pi}{2}}=0\,,\quad \lim_{x\rightarrow +\infty}r(x,y)=r_R\,,\quad \lim_{x\rightarrow +\infty}e^{+x}\partial_x r(x,y)=0\, .
\end{align}
To access the boundary conditions in the numerical scheme detailed in the following paragraph, the $x$ direction of the Riemann surface $\Sigma$ is compactified by the transformation $\xi=\tanh x$.

The partial differential equation \eqref{eq:min_cond}, for given functions $f$, $g$ and $b(r)$, is then numerically solved by means of the method of finite differences. The domain is discretized into a lattice, whilst the partial differential equation is turned into a set of coupled algebraic equations. The equations are then solved with the relaxation method in \eqref{eq:min_cond_tau}, which at this point is simply a set of first order differential equations. The numerical data are obtained using the finite difference scheme at second order as then the Neumann boundary condition take on particularly simple form, which proves sufficient to obtain numerical results within a more than acceptable accuracy. Generically, we have found it sufficient to use a grid with $\{\xi, y\}\sim \{120, 100\}$ rows each, yielding an approximation of the solution at relaxation times of the order $\tau =10^3$ with residuals in eq. \eqref{eq:residual} of order $10^{-9}$ or lower. 

\subsection{Island surfaces at zero temperature}

\begin{figure}[t!]
	\centering 
	\includegraphics[width=1\textwidth]{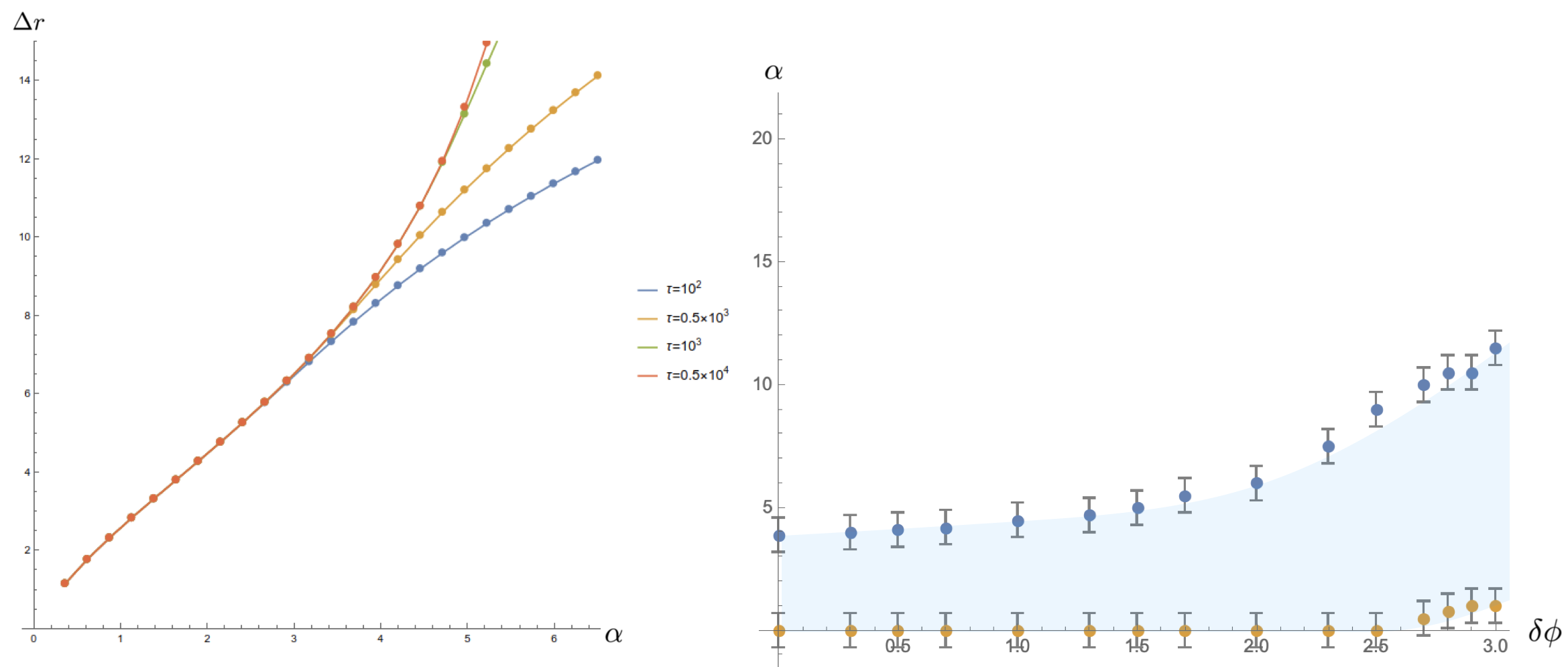}\hfill
	\caption{\small On the left: for increasing relaxation time for $\delta \phi=1.9$, $\Delta r$ becomes arbitrarily large as $\alpha$ becomes larger than a critical value. On the right: the critical geometric ratio $\alpha_{{crit}}=\sqrt{\frac{N^2}{\kappa\hat\kappa}}|_{{crit}}$ is found to increase monotonically with the dilaton variation $\delta\phi$ across the geometry. }
	\label{alphacrit-deltaphi-zeroT}
\end{figure}

Even in the absence of a black hole in the bulk geometry, the vacuum configuration yields valuable information, determining crucial features of the behavior of islands, which are relevant also at non-zero temperature. 

Zero temperature surfaces are characterized by the difference $\Delta r$ between the anchoring point in the radiation region, $r_R$, and the radial coordinate in the bag region $r_L$. For a given $\alpha$ and $\delta\phi$ and at zero temperature, the parameter $K=\sqrt{\kappa\hat\kappa}$ drops out of the extremization equation.

For $\delta\phi=0$, it was shown in \cite{Uhlemann:2021nhu} that islands cease to exist above a critical value $\alpha_{\text{crit.}}$. This can be seen numerically, also for $\delta\phi\neq0$, in the divergence of $\Delta r$. The latter, in fact, stops settling to a well-defined value. If we keep increasing the relaxation time in the solution of the PDE, once we pass a critical value of $\alpha$, the anchoring points $r_R$ and $r_L$ are pushed further apart,  as shown in Fig. \ref{alphacrit-deltaphi-zeroT}, on the left. Actually, this critical value  is drastically  affected by the non-vanishing dilaton variation. By tracking the onset of the divergence for different values of $\delta\phi$, we find that the value of the critical geometric ratio $\alpha_{\text{crit.}}$ is pushed to higher values for increasing  $\delta\phi$, as we show in Fig. \ref{alphacrit-deltaphi-zeroT}, on the right. 

This has an important consequence. First of all, for certain values of $\delta\phi\neq 0$ one can find islands at zero temperature for values of $\alpha$ which are big enough (e.g. $\alpha \sim 10$) to be in the regime of validity of \eqref{eq:grav_mass}, namely of small graviton masses. However, even on these backgrounds, one cannot adjust the graviton mass indefinitely by tuning the dilaton variation to high values, as the islands are bound to disappear for big enough values of $\delta\phi$.

We have numerically seen, in fact, that for values of $\delta\phi$ large enough $\delta\phi \gtrsim 5$, there is no more numerical convergence for any value of $\alpha$ and island surfaces cannot be found. For values $2.5<\delta\phi<5$ divergences at small values of $\alpha$ appear, even though surfaces exist for larger (but still bounded) $\alpha$, which are a signal that our code looses predictability. We thus present the numerical results for $\delta\phi\leq3$, which captures the behaviour of $\alpha_{crit}$ but still do not show significant singularities for small $\alpha$.

\begin{figure}[t!]
	\centering
	\includegraphics[width=\textwidth]{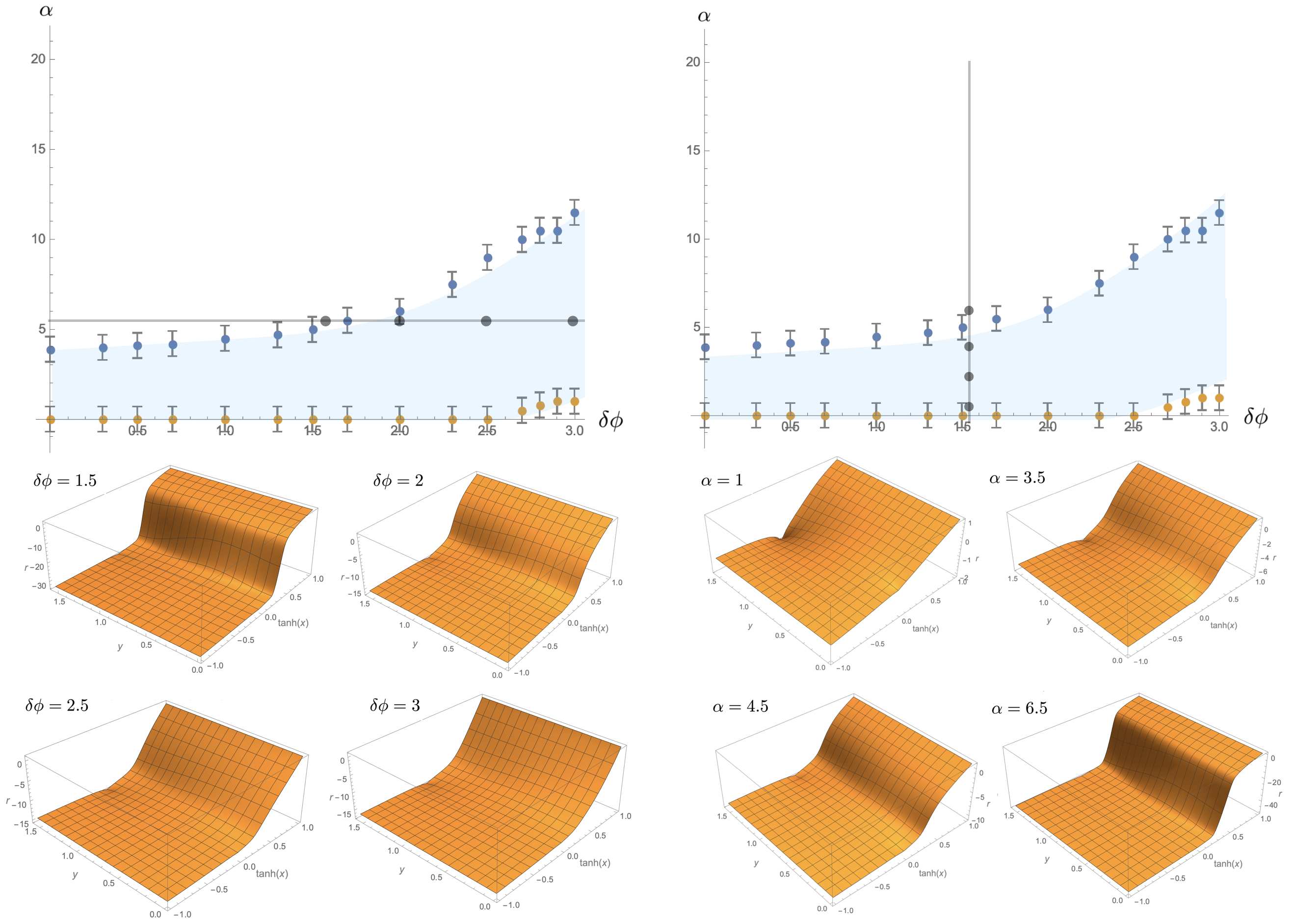}
	\caption{\small An illustration of the evolution of the minimal surface at zero temperature and non-zero dilaton variation. On the right, (a family of) backgrounds is fixed for a geometric ration $\alpha=5.5$. In both cases, the surface is indicated by a dot at the location of its values $(\delta\phi,\alpha)$ in the graph. On the left, one can see how, for a fixed dilaton variation $\delta\phi=1.7$, the surface settles into the horizon for backgrounds that approach the critical geometric ratio. At zero temperature the surfaces are shift-invariant in the radial direction, so $\Delta r$ is the meaningful quantity.
	\\[6mm]
	}\label{alphacrit-surfaces}
\end{figure}

The signal that the surfaces may develop a critical behavior can also be inferred by looking directly at the minimal surfaces, for ranging values of the parameters $(\alpha,\delta\phi)$. For a fixed value of the dilaton variation,  the islands will exist for a larger range of $\alpha$, as it is illustrated on the left of Fig. \ref{alphacrit-surfaces}, displaying how the island surfaces vary with increasing value of the geometric ratio $\alpha=\sqrt{\frac{N^2}{\kappa\hat\kappa}}$. On the other hand, increasing $\delta\phi$, in a background with fixed $\alpha$, the minimal surfaces are seen to gradually settle onto the horizon (see Fig. \ref{alphacrit-surfaces} on the right), corresponding to a larger and larger $\Delta r$. 

Summarizing, we conclude that for these type IIB uplifts, at zero temperature, in presence of a nonzero dilaton variation, quantum extremal islands can be found in a regime where the lowest-lying graviton has a small mass, but the mass cannot be tuned parametrically small by going to larger values of $\delta\phi$.

\subsection{Island surfaces at finite temperature }

At finite temperature the horizon is expected to regulate the critical effects, and islands can always exist for values of $\alpha$ above the critical value $\alpha_{\text{crit.}}$. This means one can reach a regime of small graviton mass (by appropriately tuning the dilaton variation and $\alpha$ parameters), where island surfaces can still be found. 

\begin{figure}[t!]
	\centering
	\begin{minipage}[b]{0.49\textwidth}
	\includegraphics[width=\textwidth]{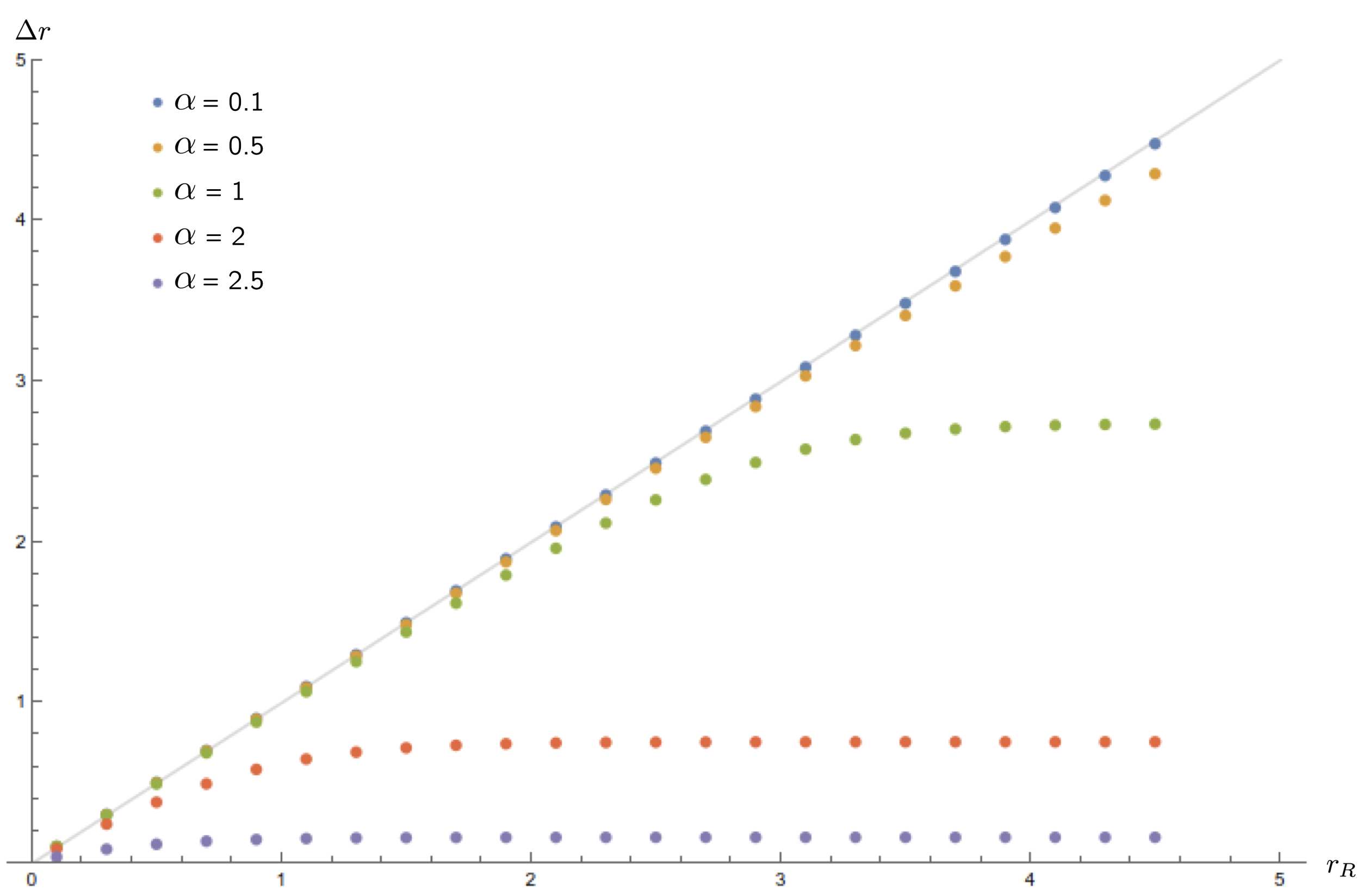}
	\end{minipage}
	\hfill
	\begin{minipage}[b]{0.49\textwidth}
	\includegraphics[width=\textwidth]{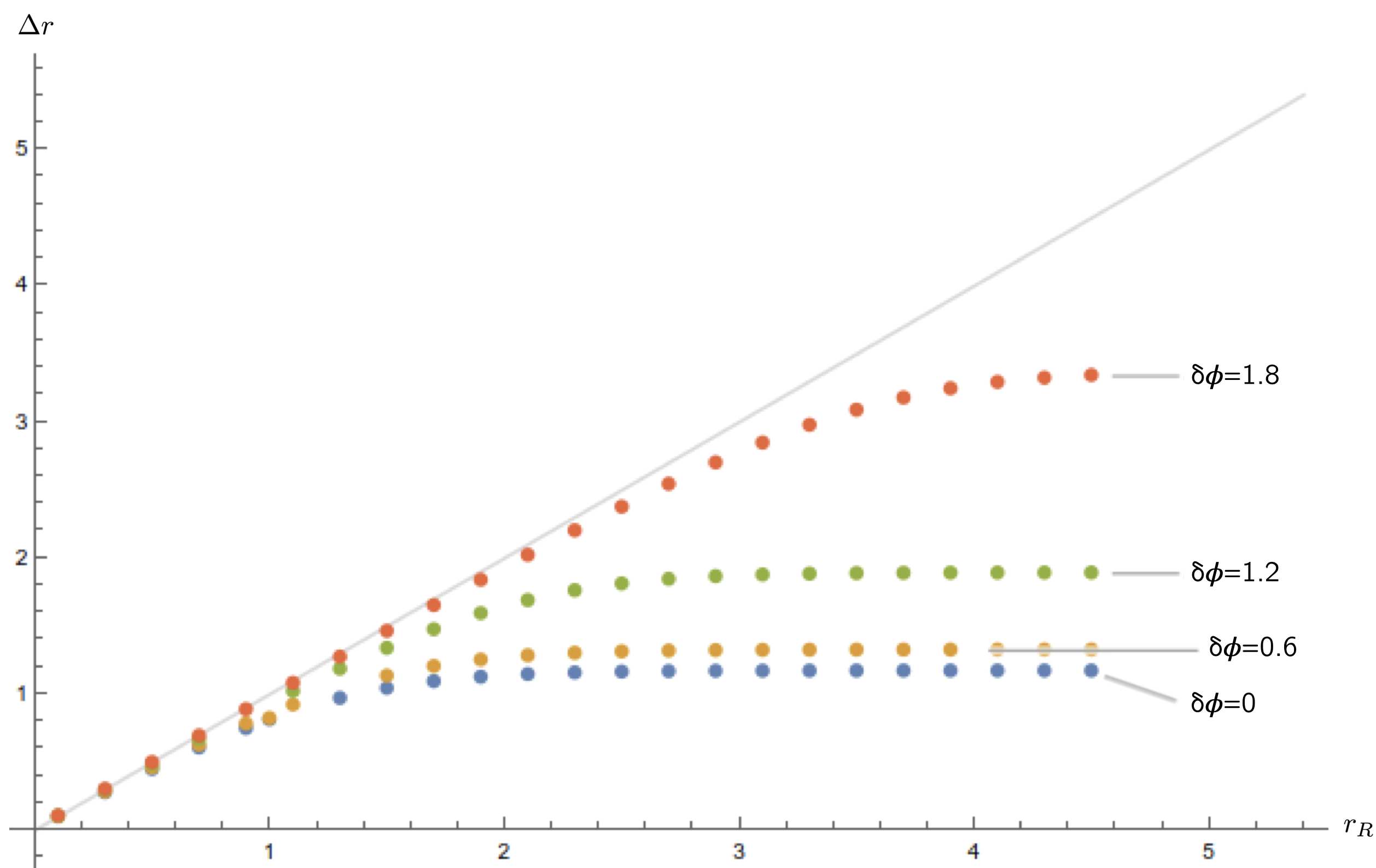}
	\end{minipage}
	\caption{\small The difference $\Delta r=r_R-r_L$ as a function of the anchoring point $r_R$ in the $AdS$ bath at finite temperature. On the left, each curve is evaluated for values of the geometric ratio $\alpha\in\{0.1,0.5,1,2,2.5\}$, with fixed $\delta\phi=0.9$. On the right, one can how the dilaton variation $\delta\phi\in\{0,0.6,1.2,1.8\}$ affects the difference $r_R-r_L$ as a function of the anchoring point $r_R$.}\label{finiteT_varalpha_Deltar}
\end{figure}
However, $\alpha_{\text{crit.}}$ still has a meaning at finite temperature, related to the behavior of the islands surface \cite{Geng:2020fxl}. More precisely, in the case of type IIB uplifts, let us look at the anchoring points of the islands surface at finite temperature. As before, $r_L$ denotes the point where the surface reaches in the $\mathrm{AdS}_4\times \mathcal M_6$ bulk, and the right radial coordinate $r_R$ denotes the location where it is anchored in the $\mathrm{AdS}_5\times S^5$ bath. In Fig. \ref{finiteT_varalpha_Deltar}, on the left, we see that at finite temperature $r_L$ is constrained to remain anchored near the horizon until the value of $\Delta r$ at zero temperature is reached, after which the difference $r_R-r_L$ remains constant,  with the same behavior noticed for $\delta\phi=0$ \cite{Uhlemann:2021nhu}. However, the constant values at which the surface settles is affected by the dilaton variation, as we show in Fig. \ref{finiteT_varalpha_Deltar}, on the right. A larger $\delta\phi$ pushes the value of $\Delta r$ to larger constants, consistent with the effect at zero temperature\footnote{The critical value $\alpha_{\text{crit.}}$, where $\Delta r$  diverges, at zero temperature, is pushed to lower values by a larger $\delta\phi$.}. As a consequence, for larger values of $\delta\phi$, $r_L$ remains longer anchored to the horizon.

The island surface stretched on the horizon will have a constant area, leading to a flat Page curve. This behavior resembles the one in the case of a gravitating bath \cite{Uhlemann:2021nhu}. Indeed, the large $\alpha$ regime corresponds to light graviton masses. One can imagine that increasing $\alpha$ will drive the background closer to the gravitating bath, where the graviton is massless. Notice that the massless graviton is a direct consequence of the compactness of the geometry, which can be obtained for any $\alpha$ by tuning $\delta\phi\to +\infty$. We can see already that our analysis suggests that these surfaces won't be the adequate objects to exhibit Page-like behavior when the bath is coupled to gravity.

\begin{figure}[th!]
	\centering
	\includegraphics[scale=0.2]{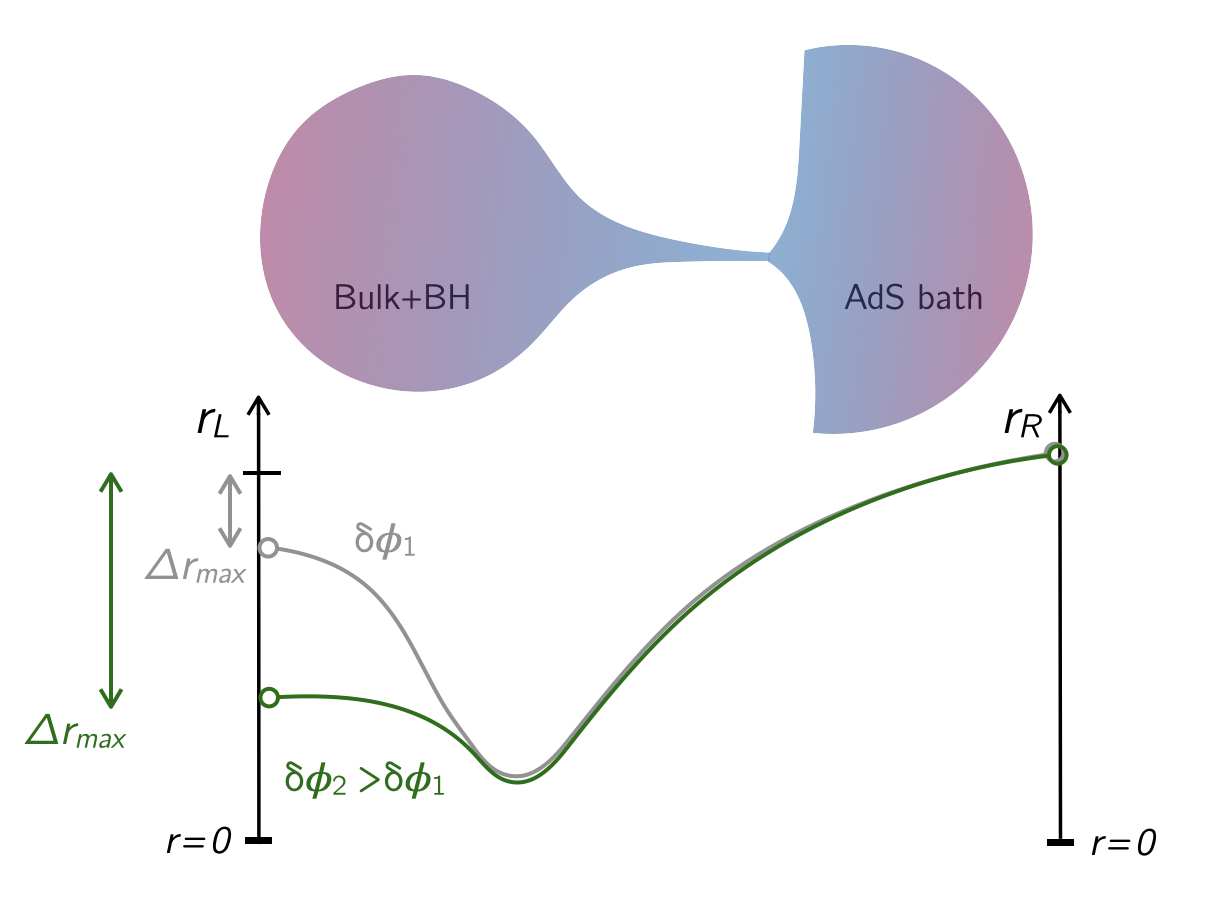}
	\caption{\small The curve illustrates the cross-section at constant $y$ of the minimal surface at a fixed value of the dilaton variation $\delta\phi=\delta\phi_2$. The value of the dilaton variation determines a minimal anchoring $r_L$ in the bulk. Increasing the dilation variation lowers this critical anchoring point for which the surface can still reach the bath. }\label{illustr_anchoring}
\end{figure}
%

%%%%%%%%%%%%%%%%%%%%%%%%%%%%%%%%%%%%%%%%
%			DISCUSSION
%%%%%%%%%%%%%%%%%%%%%%%%%%%%%%%%%%%%%%%%

\section{Conclusions and discussion}

Black hole evaporation represents an important test of our understanding of gravitational interactions. Thanks to recent developments, it is now possible to directly compute the entanglement entropy of Hawking radiation in various setups. Our work places itself within the first endeavors of this analysis in a string theory embedding. One of the main features of our analysis is to consider a string theory framework where a quantitative treatment of the graviton mass is possible, which up to now has mostly been qualitative.

In the first part of this work, we discussed the regime of validity of the massive gravity EFT, in relation with the critical value of the graviton mass relevant for the island surfaces. We have found a tension between the existence of an arbitrarily light graviton and the presence of islands below $\alpha_{\text{crit.}}$. This supports the idea that in higher dimensions it is necessary to have large enough graviton mass to correctly reproduce the Page curve \cite{Geng:2021hlu}.

In the second part of this work, we have investigated quantum extremal islands (QEI) in a 4 dimensional model of a black hole coupled to an external non-gravitating bath, uplifted to type IIB string theory, with non-constant dilaton field. 
We have considered a geometry that develops a throat of AdS$_5\times$S$^{5}$ geometry, where the mass of the lowest-lying graviton is parametrically small and can be explicitly computed. The evaporating black hole is then a system in a theory of massive gravity, that has a built-in cutoff and we have put bounds on the microscopic parameters to ensure the EFT does not break down in the region of existence of QEI surfaces, like described in the first part.

We have been able to confirm the existence of a critical value of microscopic parameters, $\alpha_{\text{crit.}}=\sqrt{\frac{N^2}{\kappa\hat\kappa}}_{\text{crit.}}$, generalizing the one in \cite{Uhlemann:2021nhu}, above which the behavior of QEI surfaces changes both at zero as well as finite temperature. By studying island surfaces in backgrounds with varying dilaton, we have shown how the critical value $\alpha_{\text{crit.}}$ is modified when increasing $\delta\phi$.

Our analysis has been able to determine that the existence of the islands does not only depend on the microscopic parameters in $\alpha$, but both on $(\alpha,\delta\phi)$, where $\delta\phi$ is the variation of the dilaton across the geometry. In particular, these two parameters determine the mass of the graviton, thus it is effectively the mass of the graviton that directly affects the behavior of the island surfaces. Our analysis has shown that there are backgrounds where the mass of the graviton can be quantitatively computed, but cannot be  tuned to arbitrarily small masses without spoiling the behavior of the island surfaces. This supports the lesson about islands and massive gravitons that comes from Karch-Randall configurations \cite{Geng:2020qvw}. 

We studied the critical parameters affecting the behavior of the island surfaces in a family of string solution with varying dilaton. In empty Anti de Sitter, one can numerically see that the surface stops existing above a critical value, although for a higher value of geometric ratio than the value at $\delta\phi=0$ (see figure \ref{alphacrit-deltaphi-zeroT}). For the black hole, our numerical study shows that, both for $\alpha$ as well as $\delta\phi$ large enough, the surfaces completely flatten on the horizon, possibly signalling a change in behavior of the island surface. In Karch-Randall setups, analogously, a critical angle determines the properties of the QEI surfaces. In that case, the behavior of the extremal surfaces in an over-critical regime is interpreted as if the surface is required to be anchored on the AdS$_4$ brane. In the type IIB construction of \cite{Uhlemann:2021nhu} for $\delta\phi=0$, and of this work for $\delta\phi\neq0$, we see that the anchoring point is stuck at the horizon, unless $r_R$ reaches far enough in the $AdS_5$ region. It also emerges, from the shape of the island surfaces at larger dilaton variation, that they relax on the horizon in approaching the critical values of the geometric parameter $\alpha$. This motivates further, future investigations to understand better what physical properties are related to this change of behavior, and to characterize the phase transitions the island surfaces undergo. It could be also interesting to discuss  the Page curve in presence of a gravitating bath in backgrounds with varying dilaton.

\subsection*{Acknowledgedments}  
We would like to thank Costas Bachas, Marine De Clerck, Ruben Monten and Christoph Uhlemann for interesting discussions and comments. The work of D.L. and I.L. is supported by the ``Origins" Excellence Cluster. This work is supported also by the German-Israeli Project (DIP) ``Holography and the Swampland".

\bibliographystyle{JHEP}
\bibliography{biblio}

\end{document}